\documentclass[%
 reprint,
 superscriptaddress,
 amsmath,amssymb,
 aps,
 pra,
]{revtex4-2}

\usepackage{graphicx}
\usepackage{dcolumn}
\usepackage{bm}

\begin{document}

\title{Absolute measurements of state-to-state rotational energy transfer between CO and H$_2$
 at interstellar temperatures}

\author{Hamza Labiad}
\affiliation{Univ Rennes, CNRS, IPR (Institut de Physique de Rennes) - UMR 6251, F-35000 Rennes, France}
\author{Martin Fournier}
\affiliation{Univ Rennes, CNRS, IPR (Institut de Physique de Rennes) - UMR 6251, F-35000 Rennes, France}
\author{Laura  A. Mertens}
\affiliation{Univ Rennes, CNRS, IPR (Institut de Physique de Rennes) - UMR 6251, F-35000 Rennes, France}
\affiliation{Division of Chemistry and Chemical Engineering, California Institute of Technology, Pasadena, California 91125, United States}
\author{Alexandre Faure}
\affiliation{Universit\'e de Grenoble Alpes, CNRS, IPAG, F-38000 Grenoble, France}
\author{David Carty}
\affiliation{Durham University, Joint Quantum Centre Durham-Newcastle, Departments of Physics and Chemistry, Lower Mountjoy, South Road, Durham DH1 3LE, UK}
\author{Thierry Stoecklin}
\affiliation{Institut des Sciences Mol\'eculaires, Universit\'e de Bordeaux, F-33405 Talence, France}
\author{Piotr Jankowski}
\affiliation{Faculty of Chemistry, Nicolaus Copernicus University in Torun, Gagarina 7, PL-87-100 Torun, Poland}
\author{Krzysztof Szalewicz}
\affiliation{Department of Physics and Astronomy, University of Delaware, Newark, DE 19716, USA}
\author{S\'ebastien D. Le Picard}
\affiliation{Univ Rennes, CNRS, IPR (Institut de Physique de Rennes) - UMR 6251, F-35000 Rennes, France}
\author{Ian R. Sims}
\email[]{ian.sims@univ-rennes1.fr}
\affiliation{Univ Rennes, CNRS, IPR (Institut de Physique de Rennes) - UMR 6251, F-35000 Rennes, France}

\date{\today}

\begin{abstract}

\noindent Experimental measurements and theoretical calculations of state-to-state rate coefficients for rotational energy transfer of CO in collision with H$_2$ are reported at the very low temperatures prevailing in dense interstellar clouds ($5-20$\,K). Detailed agreement between quantum state-selected experiments performed in cold supersonic flows using time-resolved infrared -- vacuum-ultraviolet double resonance spectroscopy and close-coupling quantum scattering calculations confirms the validity of the calculations for collisions between the two most abundant molecules in the interstellar medium.
  
\end{abstract}

\maketitle

Carbon monoxide (CO) was first identified in the interstellar medium (ISM) in 1970 towards the Orion nebula \cite{wilson70} and is the most abundant molecule in the dense ISM after molecular hydrogen (H$_{2}$). Despite its small dipole moment (0.11~D), it is the easiest molecule to detect in galaxies and is used to estimate the masses of interstellar molecular clouds, whose densest regions (so called pre-stellar cores) are the progenitors of stars. CO is also the main carbon reservoir in these ``star-forming” regions, where the temperature is $\sim$ 10~K. Emissions from CO rotational state $j = 1$ to $j = 0$ at 115 GHz and from $j = 2$ to $j = 1$ at 231 GHz are used as probes of such cold environments and, more generally, as tracers of molecular hydrogen which is invisible to radio telescopes mainly because it has no dipole moment. The dominant cooling mechanism in molecular clouds is collisional excitation of CO followed by rotational line emission, which enables the gravitational collapse of the gas that ultimately leads to the formation of stars and planetary systems.

CO molecules are not in local thermodynamic equilibrium in interstellar clouds owing to the very low hydrogen density ($\sim10^4$\,cm$^{-3}$). To understand the observed CO spectra and thus the cooling rates of molecular clouds, radiative transfer calculations of rotational state populations of CO must be made. Essential input to these calculations includes \textit{absolute} rate coefficients for the rotational energy transfer (RET) of CO in collisions with the dominant collider, H$_2$, at the temperatures found in molecular clouds ($\sim 5-20$\,K). These state-to-state rate coefficients are typically obtained by quantum scattering calculations on an ab-initio potential energy surface (PES) describing the CO-H$_2$ interaction. At low temperatures, the absolute values of these rate coefficients become especially sensitive to the details of the PES, but direct experimental measurements on an absolute scale at very low temperatures are completely lacking. Indeed, to the authors' knowledge no such measurements involving H$_2$ exist for any diatomic or polyatomic collision partner.

This letter reports direct, absolute measurements of total and state-to-state rate coefficients for the RET of CO in \emph{normal}-H$_2$ (i.e., a \emph{para-}H$_2$ to \emph{ortho-}H$_2$ ratio of 1:3) at 5.5\,K, 10\,K and 20\,K. The results are compared to state-of-the-art close-coupling quantum scattering calculations.
\begin{figure*}[htbp]
    \begin{center}
        \includegraphics[width=0.9\textwidth]{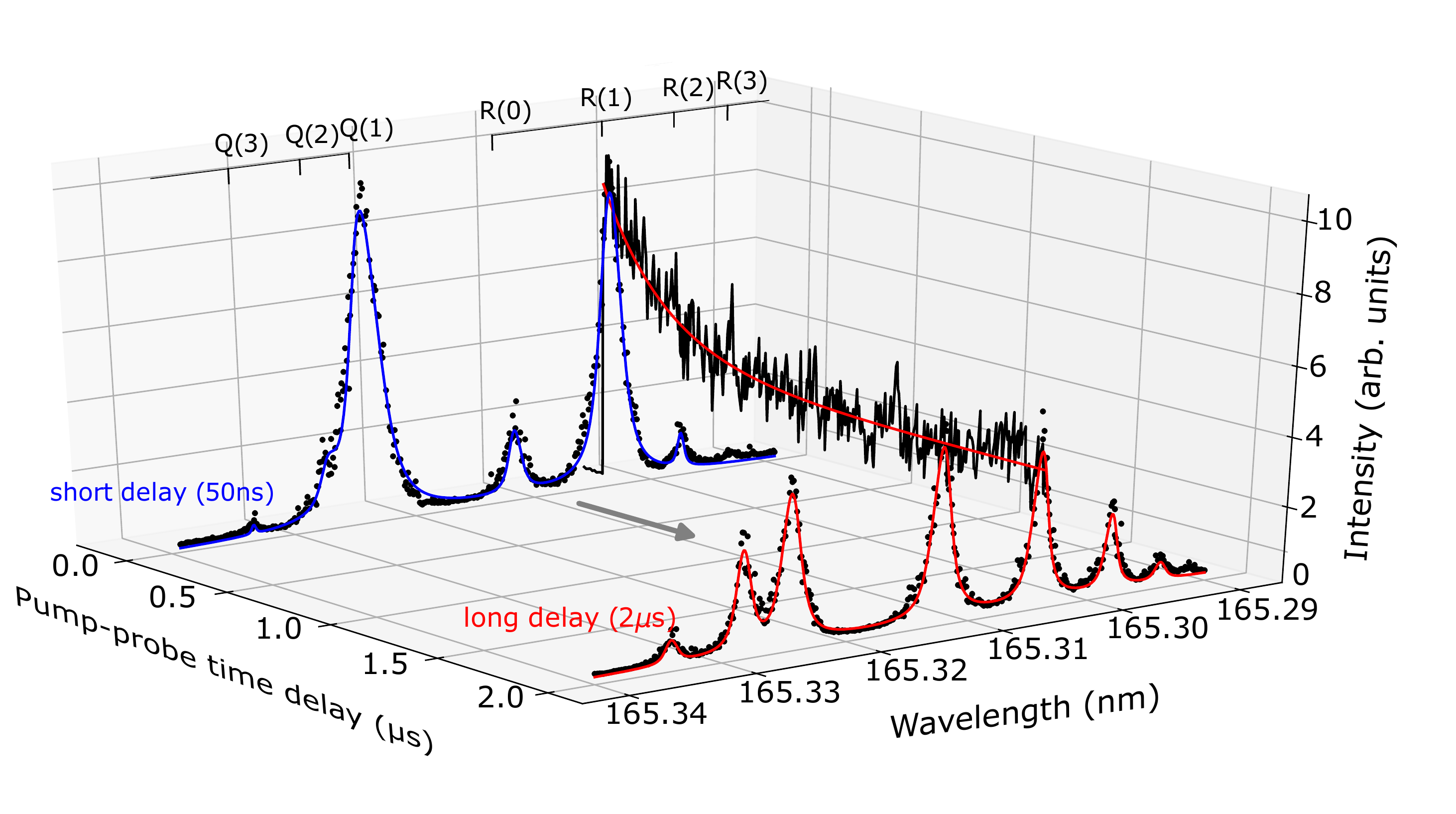}
    \end{center}
\caption{Representative IRVUVDR data from the kinetic and spectroscopic experiments at $T=5.5~$K and $[\text{H}_2]=3.2 \times 10^{16} $cm$^{-3}$. Raw data are shown in black. The short delay VUV LIF spectrum of CO($v=2$), with a fit shown in blue, was acquired at a time delay between the IR pump-laser pulse and the VUV probe-laser pulse of 50 ns after initial preparation of the $j_\mathrm{i}=1$ rotational state. The intensities of neighboring rotational lines are proportional to the state-to-state rate coefficients for RET into those levels. The spectrum at long time delay, with a fit shown in red, was acquired once thermal equilibrium of rotational states was reached, 2 $\mu$s after the pump-laser pulse. The spectra have been fitted using Voigt line profiles. The decaying signal is the intensity at the peak of the R(1) spectral line as a function of time delay, fit (in red) to a single-exponential decay.}
\label{3D_final}
\end{figure*}
\begin{figure}[htbp]
\centerline {
  \includegraphics[width=9cm]{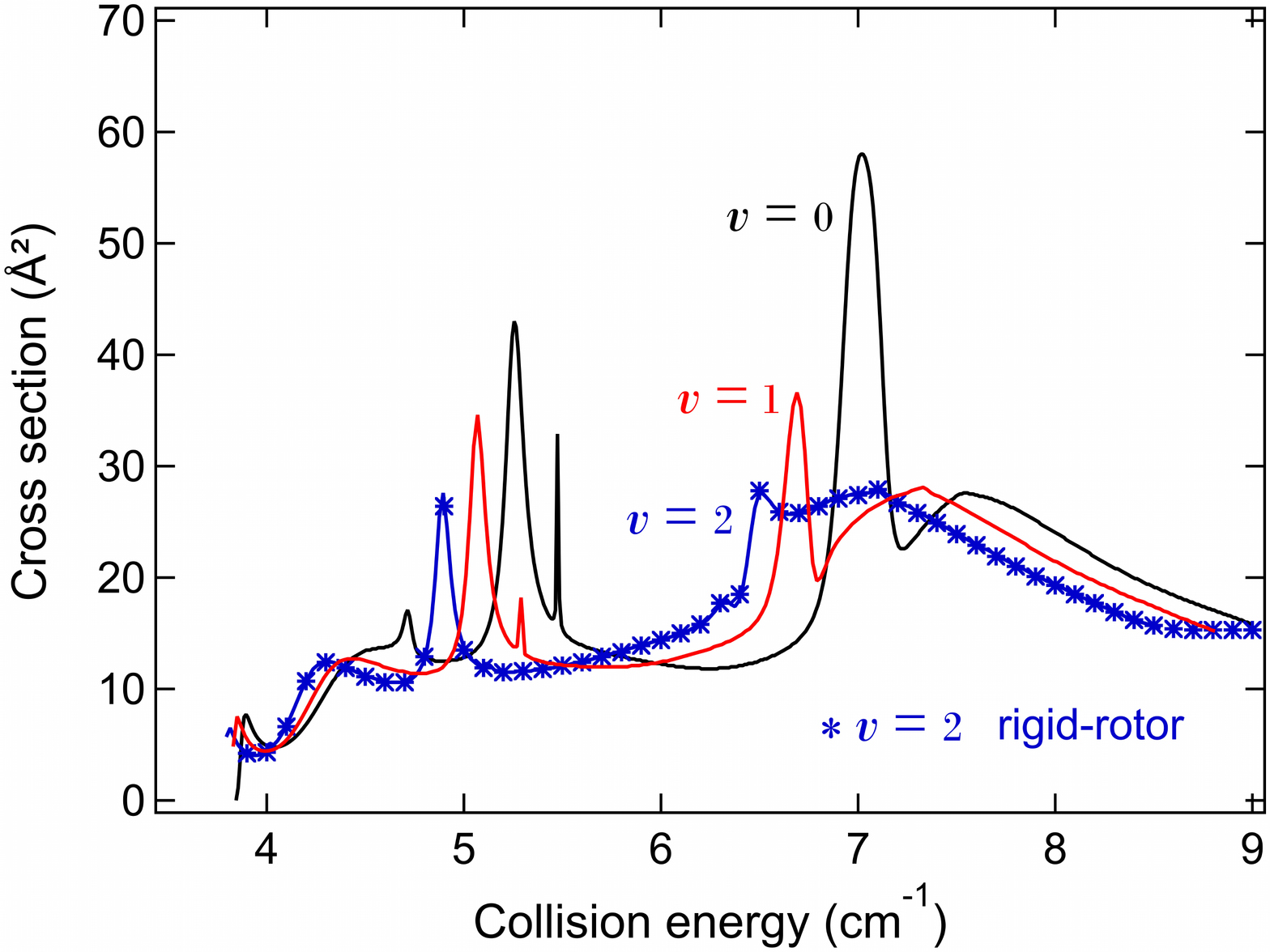}}
\caption{Calculated cross sections for the $j=0\to 1$ rotational excitation of CO in $v=0$, 1 and 2 by para-H$_2$ ($j=0$) as a function of collision energy. Solid curves correspond to scattering calculations using the 6D $V_{15}$ PES. Data points correspond to scattering calculations using the 4D $\langle V_{15} \rangle_{20}$ PES.}
\label{xs-fullD}
\end{figure}

The CO-H$_2$ collisional system has been investigated theoretically \cite{green76,jankowski98,jankowski05, yang10,jankowski12,jankowski13,chefdeville15,yang15,yang16,faure16,castro17} and experimentally \cite{brechignac80,chefdeville15,stoecklin17} in the past. The lowest temperature at which absolute rate coefficients have been measured is 77\,K \cite{brechignac80}. Recently, six-dimensional analytic CO-H$_2$
potential energy surfaces (PESs) were obtained by \citet{yang15} and by \citet{faure16}. The latter (known as $V_{15}$) was calculated using coupled cluster singles, doubles, triples and non-iterative quadruples [CCSDT(Q)] theory with very large basis sets, and was used in full-dimensional time-independent quantum scattering calculations. In contrast to \citet{yang15}, \citet{faure16} concluded that the full 6D approach was not needed at low temperature. Using a 4D treatment in which the 6D potential was averaged over ground-state monomer vibrations, an equally good agreement was obtained with the relative state-to-state cross sections measured in the crossed-beam experiments of \citet{chefdeville15}. Excellent agreement between theory and experiment was also observed in CO+D$_2$ crossed-beam experiments \cite{stoecklin17} (using an earlier $V_{12}$ potential \cite{jankowski12})   and in the measurement of CO-H$_2$ virial coefficients \cite{garberogli17}.

The current measurements were made using a CRESU (Cin\'{e}tique de R\'{e}action en Ecoulement Supersonique Uniforme or Reaction Kinetics in Uniform Supersonic Flow) apparatus \cite{CRESU,sims94}. The CRESU technique uses the supersonic expansion of a buffer gas through convergent-divergent Laval nozzles to produce uniform supersonic flows of cold gas. Each nozzle works under specific gas-pressure conditions to cool the gas to a given temperature. Special double-walled nozzles designed for H$_2$ buffer gas allow precooling to 77 K by liquid nitrogen to obtain a flow temperature as low as 5.5 K. These have only been used once before, down to 11 K by the Rennes group to study the F + H$_2$ reaction \cite{Tizniti2014}.  Further details of these special Laval nozzles and their characterization can be found in the Supplemental Material \cite{Supplemental}.

CO molecules are prepared in selected initial rovibrational states ($v=2,~j_\mathrm{i}=0,1~\mathrm{or}~4$) by absorption of pulsed infrared radiation around 2350 nm, generated by a combined optical parametric oscillator -- optical parametric amplifier (LaserVision) pumped by an injection-seeded Nd:YAG laser (Continuum Surelite EX) operating at a repetition rate of 10 Hz. Collisions with H$_2$ in the cold flow on the microsecond timescale of the experiment cause population transfer to final rovibrational states ($v=2,~j_\mathrm{f}$). These states are probed by pulsed laser-induced fluorescence (LIF) via the $\mathrm{A}\,^1\Pi(v=0)\leftarrow\mathrm{X}\,^1\Sigma^+(v=2)$ vacuum ultraviolet (VUV) vibronic band at 166\,nm. Further details of this so-called infrared -- vacuum ultraviolet double resonance (IRVUVDR) technique \cite{carty_rotational_2004,mertens_rotational}, along with a schematic diagram of the experimental arrangement, may be found in the Supplemental Material. Using this IRVUVDR technique, two types of experiment are possible, \textit{kinetic} and \textit{spectroscopic}.

In \textit{kinetic} experiments, an initially prepared $j_\mathrm{i}$ state is probed and the LIF intensity is measured as a function of time, as shown in Fig.~\ref{3D_final}. The rate coefficient for total RET from $j_\mathrm{i}$ to all $j_\mathrm{f}$ states, $k_\mathrm{tot}$, is extracted from the resulting exponential decay, taking into account transfer back into the probed state as described in previous work \cite{james_combined_1998,carty_rotational_2004,mertens_rotational}. 

In \textit{spectroscopic} experiments, an initial $j_\mathrm{i}$ state is prepared and the wavelength of the VUV laser is scanned to acquire a spectrum that probes all possible $j_\mathrm{f}$ states. As shown in Fig.~\ref{3D_final}, two spectra are obtained, one at 15--50\,ns after the IR pulse, when only a small amount of RET has occurred, and one at 2--3\,$\mu$s when thermal equilibrium of rotational state populations has been established. Voigt lineshapes are fitted to each rotational line in both spectra to extract line intensities, which are then used to obtain state-to-state rate coefficients according to the formula:\\
\begin{equation}
   k_{j_i \rightarrow j_f} = \left( \frac{I_{j_f}^{\delta t} f_{j_f}^{eq} }{I_{j_f}^{eq}}  \right) \frac{1}{\delta t[\text{H}_2]}
\end{equation}
where $I_{j_f}^{\delta t}$ is the line intensity corresponding to the final rotational state $j_f$ at short time delay $\delta t$, $f_{j_f}^{eq}$ is the Boltzmann factor (fractional population at rotational equilibrium)  for $j_f$, $I_{j_f}^{eq}$ is the line intensity for $j_f$ at equilibrium, and $[$H$_2]$ is the H$_2$ concentration. The validity of this approach has been discussed in previous work \cite{james_combined_1998,carty_rotational_2004,mertens_rotational}, but briefly, it relies on the short time delay being small enough that transfer via multiple levels may be neglected, while the long time delay is sufficient to ensure full rotational relaxation, but still short enough to avoid significant vibrational relaxation. The latter is assured as vibrational relaxation in this system is many orders of magnitude slower than rotational relaxation \cite{yang15}. CO-CO self-relaxation can be neglected as [CO]/[H$_2$] is maintained at $\lesssim$0.05.

\begin{figure*}[htbp]
\centerline {
\includegraphics[width=18.5cm]{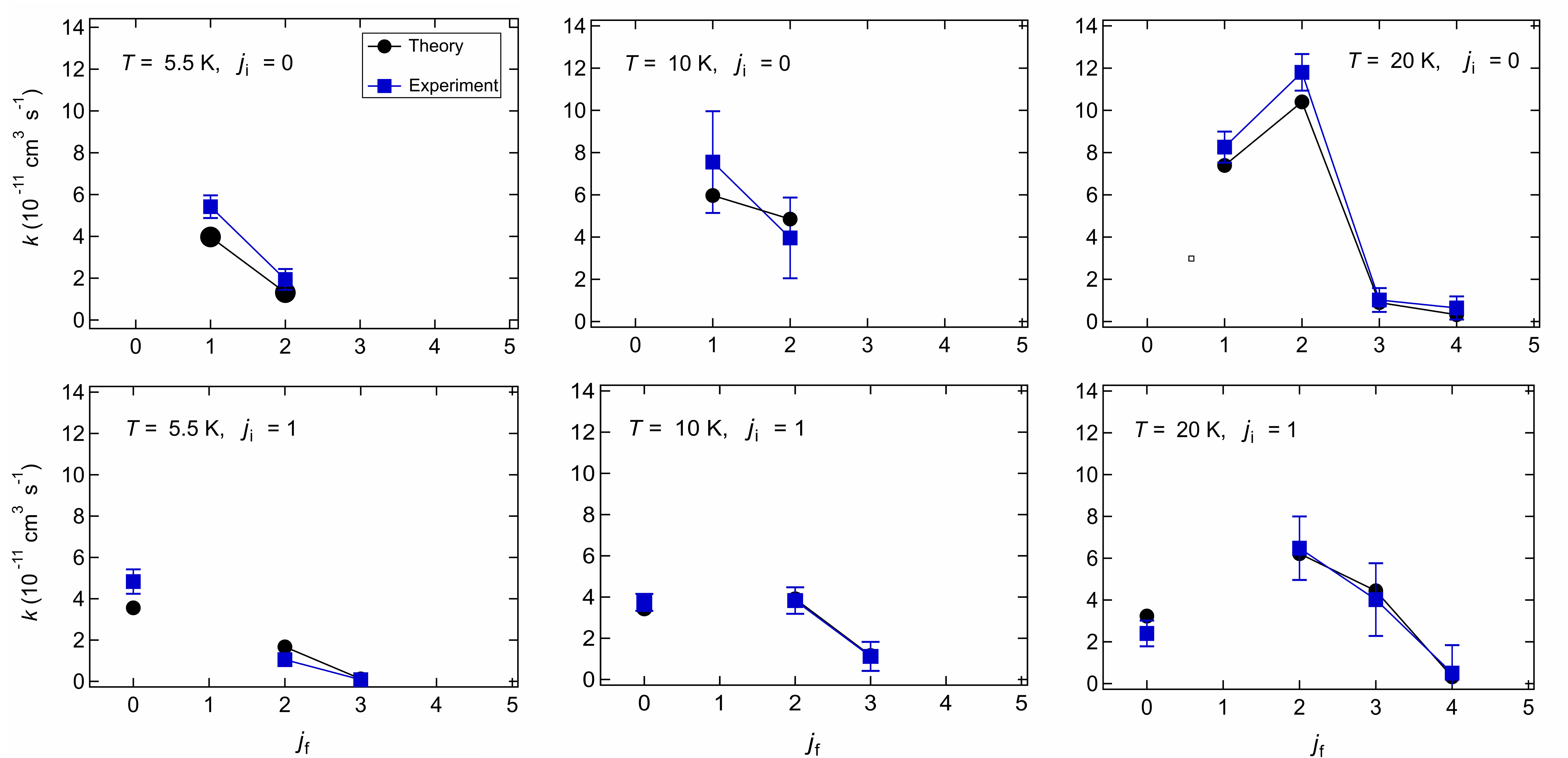}}
\caption{Measured state-to-state rate coefficients (squares) and theoretical state-to-state rate coefficients (circles), calculated using the 4D $\langle V_{15} \rangle_{20}$ PES for $j_\mathrm{i}=0\rightarrow j_\mathrm{f}$ and $j_\mathrm{i}=1\rightarrow j_\mathrm{f}$ at 5.5\,K, 10\,K and 20\,K. Error bars correspond to 2$\sigma$.}
\label{All_st}
\end{figure*}

\begin{table}
\begin{center}
\begin{tabular}{|c||c|c|c|}

\multicolumn{4}{c}{\textbf{\textit{T} = 20 K}} \\ 

\hline
 & \multicolumn{3}{c|}{$j_\mathrm{i}$} \\ 
\hline
$j_\mathrm{f}$ & 0 & 1 & 4 \\
\hline
0 & - & $2.4\pm0.6$ (3.2)
 & $1.0\pm0.8$ (0.6)
 \\ 

1 & $8.3\pm0.7$ (7.4)
 & - & $2.2\pm1.2$ (1.3)
 \\ 

2 & $11.8\pm0.9$ (10.3)
 & $6.5\pm1.5$ (6.2)
 & $8.5\pm1.1$ (8.4)
 \\ 

3 & $1.0\pm0.6$ (0.9)
 & $4.0\pm1.7$ (4.4)
 & $7.9\pm1.1$ (7.5)
 \\ 

4 & $0.6\pm0.5$ (0.03)
 & $0.5\pm1.3$ (0.3)
 & - \\ 

5 & (0.03) & (0.08) & $1.7\pm0.7$ (2.3)
 \\
\hline
$\sum k$ & $21.7\pm1.4$ (18.7) & $13.4\pm2.7$ (14.2) & $21.3\pm2.2$ (20.1) \\
\hline
$k_\mathrm{tot}$ & $16.1\pm4.0$ & $13.4\pm2.6$ & $20.1\pm4.1$ \\
\hline

\multicolumn{4}{c}{\textbf{\textit{T} = 10 K}} \\

\hline
$j_\mathrm{f}$ & 0 & 1 & 4 \\
\hline
0 & - & $3.7\pm0.4$ (3.4) & (0.5)
 \\ 

1 & $7.6\pm2.4$ (6.0)
 & - & (1.1) \\ 

2 & $4.0\pm1.9$ (4.9)
 & $3.8\pm0.6$ (3.9) & (8.3)
 \\ 

3 & (0.02) & $1.1\pm0.7$ (1.2) & (8.1)
 \\
 \hline
$\sum k$ & $11.6\pm3.1$ (11.0) & $8.6\pm1.0$ (8.5) & (18.0) \\
\hline
$k_\mathrm{tot}$ & $6.6\pm0.9$ & $7.9\pm0.9$ & - \\
\hline

\multicolumn{4}{c}{\textbf{\textit{T} = 5.5 K}} \\ 

\hline
$j_\mathrm{f}$ & 0 & 1 & 4\\
\hline
0 & - & $4.8\pm0.6$ (3.6) & (0.5)
 \\
1 & $5.4\pm0.5$ (4.0)
 & - & (1.0) \\ 

2 & $1.9\pm0.5$ (1.3)
 & $1.1\pm0.3$ (1.7) & (7.8)
 \\ 

3 & (0.01) & $0.08\pm0.08$ (0.1) & (8.5)
 \\
\hline
$\sum k$ & $7.3\pm0.7$ (5.3) & $6.0\pm0.7$ (5.4) & (17.8) \\ 
\hline $k_\mathrm{tot}$ & $5.8\pm1.1$ & $6.0\pm0.7$ & - \\ 
\hline %
\end{tabular} 
\caption{Measured state-to-state rate coefficients and theoretical values (in parentheses) for RET in CO -- H$_2$ collisions, calculated using the 4D $\langle V_{15} \rangle_{20}$ PES, for $j_\mathrm{i}=0, 1, 4\rightarrow j_\mathrm{f}$ at 5.5\,K, 10\,K and 20\,K. Units are in 10$^{-11}$\,cm$^3$\,s$^{-1}$ and error limits correspond to 2$\sigma$ statistical errors. The sum of the state-to-state rate coefficients, $\Sigma k,$ and the equivalent rate coefficients, $k_\mathrm{tot}$, from kinetic experiments are also given.} 
\label{st_table}%
\end{center}

\end{table}%

\begin{figure}
\centerline {
\includegraphics[width=8cm]{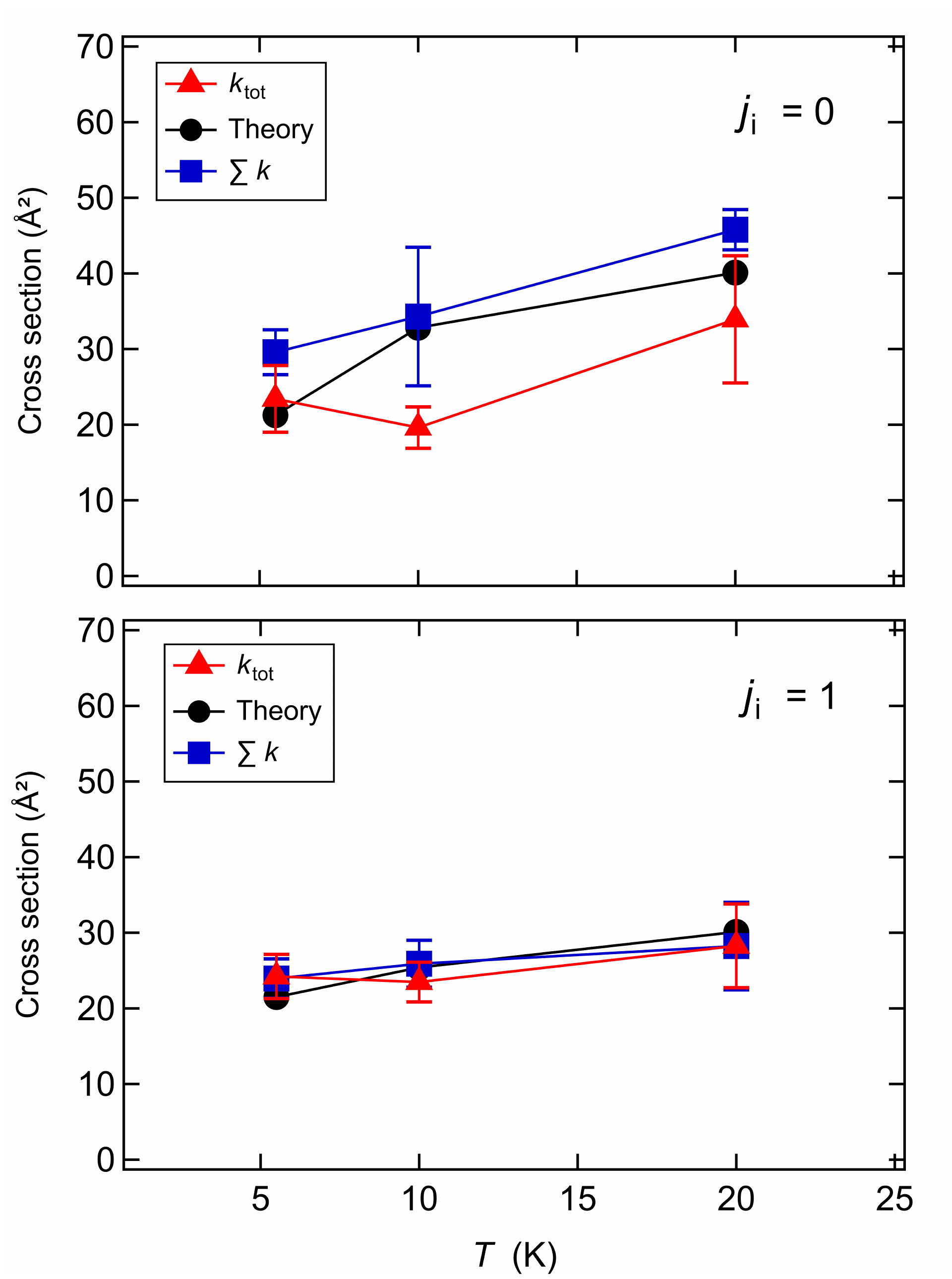}}
\caption{Temperature dependence of measured and theoretical thermally-averaged cross-sections for $j_\mathrm{i}$=0 and $j_\mathrm{i}$=1 to all $j_\mathrm{f}$ states calculated from $\Sigma k$ and from $k_\mathrm{tot}$.}
\label{Cross_section_T}
\end{figure}

Full-dimensional scattering calculations were performed with the \texttt{DIDIMAT} code \cite{faure16} using the 6D $V_{15}$ PES for the $v=0$, 1 and 2 states of CO. These were compared to close-coupling calculations performed using the \texttt{MOLSCAT} \cite{molscat} scattering code. In the latter, the rigid-rotor approximation was applied and the 6D $V_{15}$ PES was averaged over the $v=2, j=0$ vibrational wave function of CO and the $v=0, j=0$ vibrational wave function of H$_2$ to give a 4D PES, $\langle V_{15} \rangle_{20}$. In both types of calculations, the (time-independent) close-coupling approach was used to solve a set of coupled radial equations derived from the Schrödinger equation. In the rigid-rotor equations, the rotational constants for CO and H$_2$ were fixed at their experimental values $B_2(\rm CO)$=1.8875~cm$^{-1}$ and $B_0(H_2)=59.322$~cm$^{-1}$ \cite{Huber:79}. Convergence of calculations was carefully checked with respect to the size of the basis sets and the propagation parameters. The rotational basis set of CO thus included levels $j=0-20$ while that of H$_2$ included $j_{\rm H_2}=0, 2$ for para-H$_2$ and $j_{\rm H_2}=1, 3$ for ortho-H$_2$. Cross sections $\sigma_{j_i \to j_f}(E_c)$ were computed as function of the collision energy $E_c$ from the appropriate T-matrices, as described in \citet{stoecklin17}.

Fig.~\ref{xs-fullD} shows cross sections as a function of collision energy for $j=0\rightarrow 1$ RET of CO upon collision with \emph{para}-H$_2$($v=0,j=0$). The cross sections resulting from full-dimensional scattering calculations for CO($v=0$, 1 and 2) are shown as solid lines, while the data points result from the approximate rigid-rotor (4D) close-coupling calculations for CO($v=2$). They agree extremely well with the full-dimensional scattering calculations. It can also be seen that while the positions and magnitudes of resonance peaks differ for different vibrational states, the background cross-sections are similar in magnitude. Therefore, while it is relevant to perform calculations for the vibrational state populated in the experiments (especially at the lowest temperature), we note that the shift and decrease of resonance peaks will affect the rate coefficient values by a few percent only. In the following, only the rigid-rotor 4D calculations are presented and discussed. The rotational rate coefficients were computed by averaging the corresponding cross sections over Maxwell-Boltzmann velocity distributions at 5.5\,K, 10\,K and 20\,K.

 Table~\ref{st_table} and Fig.~\ref{All_st} compare measured state-to-state rate coefficients and theoretical values (calculated using the $\langle V_{15} \rangle_{20}$ PES) for $j_\mathrm{i}=0\rightarrow j_\mathrm{f}$ and $j_\mathrm{i}=1\rightarrow j_\mathrm{f}$ at 5.5\,K, 10\,K and 20\,K. Table~\ref{st_table} also displays values for $j_\mathrm{i}=4\rightarrow j_\mathrm{f}$ calculated at 5.5\,K, 10\,K and 20\,K and measured at 20 K. At 5.5\,K and at 10\,K, $j_\mathrm{i}=4$ was not accessible in the rovibrational state preparation as rotational states above $j=2$ are not populated appreciably at such low temperatures. The agreement between experiment and theory is excellent and the results are all self-consistent within the principle of detailed balance \cite{[{i.e. ${{k}_{1\to 0}}={{k}_{0\to 1}}\left( {{{g}_{1}}}/{{{g}_{0}}}\; \right)\exp \left( {-\Delta {{E}_{0-1}}}/{kT}\; \right)$ where $k_{0 \rightarrow 1}$ and $k_{1 \rightarrow 0}$ are the forward and backward rate coefficients between levels 0 and 1, ${{g}_{0}}$ and ${{g}_{1}}$ are the degeneracies associated with those levels and ${\Delta {E}_{0-1}}$ is the difference in energy. }]Mahan1974}. Table~\ref{st_table} also contains measured total RET rate coefficients, $k_\mathrm{tot}$, and the sum of the state-to-state rate coefficients, $\Sigma k$; these agree within experimental error, with the marginal exception of $j_i=0$ at 10\,K.

Fig.~\ref{Cross_section_T} shows a plot of thermally-averaged cross sections for RET from $j_\mathrm{i}=0$ and 1 to all $j_\mathrm{f}$ states at all temperatures measured. The cross sections were obtained by dividing the measured and theoretical $\Sigma k$ values, for a given $j_\mathrm{i}$ and temperature, by the mean thermal velocity at that temperature. The same was done using measured $k_\mathrm{tot}$ values from the kinetic experiments. The cross sections tend to decrease with decreasing temperature. As the temperature, and thus the relative velocity between the CO and H$_2$, decreases, the probability of a collision occurring increases as the long-range van der Waals attraction brings the molecules together. This promotes an increase in cross section as the temperature is lowered, so a second mechanism that results in a decrease must dominate. To understand this, consider the state-to-state rate coefficients again. Positive $\Delta j$ collisions are endothermic, whereas negative $\Delta j$ collisions are exothermic. For $j_\mathrm{i}=0$, RET is purely endothermic. It can be seen in the top row of Fig.~\ref{All_st} that the state-to-state rate coefficients for $\Delta j=+1$ increase with increasing temperature as the proportion of collisions that are energetic enough to cause RET increases. The same is true for $\Delta j=+2$ RET. Only at 20\,K is the proportion of collisions energetic enough to cause $\Delta j=+3$ and $\Delta j=+4$ RET. For $j_\mathrm{i}=1$, the rate coefficients for $\Delta j=-1$ RET are almost constant with temperature due to this process being exothermic. For the endothermic, positive $\Delta j$ RET from $j_\mathrm{i}=1$, the trend is similar to that for $j_\mathrm{i}=0$, but the evidence for a $\Delta j=+2$ propensity is weak owing to a suppression of $\Delta j>1$ RET as a result of the greater spacing of rotational energy levels at higher $j$. The net result is that there is an increase in $\sum k$ (which equals $k_\mathrm{tot}$), and hence cross section, as the temperature increases that more than cancels out the opposite temperature dependence due to the long-range van der Waals attraction.


At 20 K, it can also be seen that the rate coefficient for $j_\mathrm{i}=0\rightarrow j_\mathrm{f}=2$ is larger than that for $j_\mathrm{i}=0\rightarrow j_\mathrm{f}=1$. Such a $\Delta j$ propensity is due to quantum mechanical effects in scattering involving ``almost
homonuclear" molecules and is related to the anisotropy of the potential. It can be understood in terms of the partial-wave expansion of the potential $V$ where for a homonuclear diatomic molecule the term $l = 1$ is missing, so the matrix element of $V$ between the initial $j_{\rm i}$ function and the final $j_{\rm i} + 1$ function is zero. This selection rule, $\Delta j=$ even, becomes only a propensity rule for an ``almost homonuclear'' molecule like CO. This effect was first investigated by \citet{mccurdy77} in the context of classical $S$-matrix theory.

In conclusion, absolute state-to-state rate coefficients and thermally-averaged cross-sections for rotational energy transfer of CO in collision with \textit{normal}-H$_2$ have been measured directly using infrared vacuum ultraviolet double resonance spectroscopy in a CRESU apparatus at interstellar temperatures. The CO states measured were $v=2,j_\mathrm{i}=0,1$ and 4 at 20\,K and, $v=2,j_\mathrm{i}=0$ and 1 at 10\,K and at 5.5\,K. The excellent and detailed agreement on an absolute basis between direct experimental measurements and theoretical calculations using both the 6D $V_{15}$ PES and the approximate, 4D $\langle V_{15}\rangle_{20}$ PES provides rare benchmarking of theory. In particular, we believe that the extensive set of collisional data computed by \citet{yang10} and based on the previous $V_{04}$ PES of Jankowski and Szalewicz \cite{jankowski05} can be employed reliably. Indeed, as shown by \citet{faure16}, the difference between $V_{15}$ and $V_{04}$ has only a  moderate impact on the low-energy integral cross sections, and even less so on the rate coefficients (see also the discussion in \citet{chefdeville15}). The calculations can now be used with complete confidence in the interpretation of spectra of astronomical CO in star-forming molecular clouds and in the understanding of molecular-cloud cooling that leads to star formation by gravitational collapse.

This work was supported by the Agence Nationale de la Recherche (ANR-HYDRIDES), contract ANR-12-BS05-0011-01, by the R\'egion de Bretagne and the French National Programme ``Physique et Chimie du Milieu Interstellaire'' (PCMI) of CNRS/INSU with INC/INP co-funded by CEA and CNES. DC would like to thank the Universit\'{e} de Rennes 1 for support via the Visiting Professor programme. This work was also supported in part by the National Science Foundation under Grant No. CHE-1900551. PJ acknowledges support from Polish National Science Centre grant no. 2017/25/B/ST4/01300.

\bibliography{coh2}

\begin{thebibliography}{31}%
\makeatletter
\providecommand \@ifxundefined [1]{%
 \@ifx{#1\undefined}
}%
\providecommand \@ifnum [1]{%
 \ifnum #1\expandafter \@firstoftwo
 \else \expandafter \@secondoftwo
 \fi
}%
\providecommand \@ifx [1]{%
 \ifx #1\expandafter \@firstoftwo
 \else \expandafter \@secondoftwo
 \fi
}%
\providecommand \natexlab [1]{#1}%
\providecommand \enquote  [1]{``#1''}%
\providecommand \bibnamefont  [1]{#1}%
\providecommand \bibfnamefont [1]{#1}%
\providecommand \citenamefont [1]{#1}%
\providecommand \href@noop [0]{\@secondoftwo}%
\providecommand \href [0]{\begingroup \@sanitize@url \@href}%
\providecommand \@href[1]{\@@startlink{#1}\@@href}%
\providecommand \@@href[1]{\endgroup#1\@@endlink}%
\providecommand \@sanitize@url [0]{\catcode `\\12\catcode `\$12\catcode
  `\&12\catcode `\#12\catcode `\^12\catcode `\_12\catcode `\%12\relax}%
\providecommand \@@startlink[1]{}%
\providecommand \@@endlink[0]{}%
\providecommand \url  [0]{\begingroup\@sanitize@url \@url }%
\providecommand \@url [1]{\endgroup\@href {#1}{\urlprefix }}%
\providecommand \urlprefix  [0]{URL }%
\providecommand \Eprint [0]{\href }%
\providecommand \doibase [0]{https://doi.org/}%
\providecommand \selectlanguage [0]{\@gobble}%
\providecommand \bibinfo  [0]{\@secondoftwo}%
\providecommand \bibfield  [0]{\@secondoftwo}%
\providecommand \translation [1]{[#1]}%
\providecommand \BibitemOpen [0]{}%
\providecommand \bibitemStop [0]{}%
\providecommand \bibitemNoStop [0]{.\EOS\space}%
\providecommand \EOS [0]{\spacefactor3000\relax}%
\providecommand \BibitemShut  [1]{\csname bibitem#1\endcsname}%
\let\auto@bib@innerbib\@empty
\bibitem [{\citenamefont {{Wilson}}\ \emph {et~al.}(1970)\citenamefont
  {{Wilson}}, \citenamefont {{Jefferts}},\ and\ \citenamefont
  {{Penzias}}}]{wilson70}%
  \BibitemOpen
  \bibfield  {author} {\bibinfo {author} {\bibfnamefont {R.~W.}\ \bibnamefont
  {{Wilson}}}, \bibinfo {author} {\bibfnamefont {K.~B.}\ \bibnamefont
  {{Jefferts}}},\ and\ \bibinfo {author} {\bibfnamefont {A.~A.}\ \bibnamefont
  {{Penzias}}},\ }\bibfield  {title} {\bibinfo {title} {Carbon monoxide in the
  {Orion Nebula}},\ }\href {https://doi.org/10.1086/180567} {\bibfield
  {journal} {\bibinfo  {journal} {\apj}\ }\textbf {\bibinfo {volume} {161}},\
  \bibinfo {pages} {L43} (\bibinfo {year} {1970})}\BibitemShut {NoStop}%
\bibitem [{\citenamefont {{Green}}\ and\ \citenamefont
  {{Thaddeus}}(1976)}]{green76}%
  \BibitemOpen
  \bibfield  {author} {\bibinfo {author} {\bibfnamefont {S.}~\bibnamefont
  {{Green}}}\ and\ \bibinfo {author} {\bibfnamefont {P.}~\bibnamefont
  {{Thaddeus}}},\ }\bibfield  {title} {\bibinfo {title} {{Rotational excitation
  of CO by collisions with He, H, and H$_2$ under conditions in interstellar
  clouds}},\ }\href {https://doi.org/10.1086/154333} {\bibfield  {journal}
  {\bibinfo  {journal} {\apj}\ }\textbf {\bibinfo {volume} {205}},\ \bibinfo
  {pages} {766} (\bibinfo {year} {1976})}\BibitemShut {NoStop}%
\bibitem [{\citenamefont {Jankowski}\ and\ \citenamefont
  {Szalewicz}(1998)}]{jankowski98}%
  \BibitemOpen
  \bibfield  {author} {\bibinfo {author} {\bibfnamefont {P.}~\bibnamefont
  {Jankowski}}\ and\ \bibinfo {author} {\bibfnamefont {K.}~\bibnamefont
  {Szalewicz}},\ }\bibfield  {title} {\bibinfo {title} {{\it Ab initio}
  potential energy surface and infrared spectra of {H$_2$--CO} and {D$_2$--CO}
  van der {Waals} complexes},\ }\href {https://doi.org/10.1063/1.475347}
  {\bibfield  {journal} {\bibinfo  {journal} {\jcp}\ }\textbf {\bibinfo
  {volume} {108}},\ \bibinfo {pages} {3554} (\bibinfo {year}
  {1998})}\BibitemShut {NoStop}%
\bibitem [{\citenamefont {Jankowski}\ and\ \citenamefont
  {Szalewicz}(2005)}]{jankowski05}%
  \BibitemOpen
  \bibfield  {author} {\bibinfo {author} {\bibfnamefont {P.}~\bibnamefont
  {Jankowski}}\ and\ \bibinfo {author} {\bibfnamefont {K.}~\bibnamefont
  {Szalewicz}},\ }\bibfield  {title} {\bibinfo {title} {A new {\it ab initio}
  interaction energy surface and high-resolution spectra of the {H$_2$--CO} van
  der {Waals} complex},\ }\href {https://doi.org/10.1063/1.2008216} {\bibfield
  {journal} {\bibinfo  {journal} {\jcp}\ }\textbf {\bibinfo {volume} {123}},\
  \bibinfo {pages} {104301} (\bibinfo {year} {2005})}\BibitemShut {NoStop}%
\bibitem [{\citenamefont {Yang}\ \emph {et~al.}(2010)\citenamefont {Yang},
  \citenamefont {Stancil}, \citenamefont {Balakrishnan},\ and\ \citenamefont
  {Forrey}}]{yang10}%
  \BibitemOpen
  \bibfield  {author} {\bibinfo {author} {\bibfnamefont {B.~H.}\ \bibnamefont
  {Yang}}, \bibinfo {author} {\bibfnamefont {P.~C.}\ \bibnamefont {Stancil}},
  \bibinfo {author} {\bibfnamefont {N.}~\bibnamefont {Balakrishnan}},\ and\
  \bibinfo {author} {\bibfnamefont {R.~C.}\ \bibnamefont {Forrey}},\ }\bibfield
   {title} {\bibinfo {title} {Rotational quenching of {CO} due to {H$_2$}
  collisions},\ }\href {https://doi.org/10.1088/0004-637x/718/2/1062}
  {\bibfield  {journal} {\bibinfo  {journal} {\apj}\ }\textbf {\bibinfo
  {volume} {718}},\ \bibinfo {pages} {1062} (\bibinfo {year}
  {2010})}\BibitemShut {NoStop}%
\bibitem [{\citenamefont {Jankowski}\ \emph {et~al.}(2012)\citenamefont
  {Jankowski}, \citenamefont {McKellar},\ and\ \citenamefont
  {Szalewicz}}]{jankowski12}%
  \BibitemOpen
  \bibfield  {author} {\bibinfo {author} {\bibfnamefont {P.}~\bibnamefont
  {Jankowski}}, \bibinfo {author} {\bibfnamefont {A.~R.~W.}\ \bibnamefont
  {McKellar}},\ and\ \bibinfo {author} {\bibfnamefont {K.}~\bibnamefont
  {Szalewicz}},\ }\bibfield  {title} {\bibinfo {title} {Theory untangles the
  high-resolution infrared spectrum of the {\em ortho} {H$_2$-CO} van der
  {Waals} complex},\ }\href {https://doi.org/10.1126/science.1221000}
  {\bibfield  {journal} {\bibinfo  {journal} {Science}\ }\textbf {\bibinfo
  {volume} {336}},\ \bibinfo {pages} {1147} (\bibinfo {year}
  {2012})}\BibitemShut {NoStop}%
\bibitem [{\citenamefont {Jankowski}\ \emph {et~al.}(2013)\citenamefont
  {Jankowski}, \citenamefont {Surin}, \citenamefont {Potapov}, \citenamefont
  {Schlemmer}, \citenamefont {McKellar},\ and\ \citenamefont
  {Szalewicz}}]{jankowski13}%
  \BibitemOpen
  \bibfield  {author} {\bibinfo {author} {\bibfnamefont {P.}~\bibnamefont
  {Jankowski}}, \bibinfo {author} {\bibfnamefont {L.~A.}\ \bibnamefont
  {Surin}}, \bibinfo {author} {\bibfnamefont {A.~V.}\ \bibnamefont {Potapov}},
  \bibinfo {author} {\bibfnamefont {S.}~\bibnamefont {Schlemmer}}, \bibinfo
  {author} {\bibfnamefont {A.~R.~W.}\ \bibnamefont {McKellar}},\ and\ \bibinfo
  {author} {\bibfnamefont {K.}~\bibnamefont {Szalewicz}},\ }\bibfield  {title}
  {\bibinfo {title} {A comprehensive experimental and theoretical study of
  {H$_2$-CO} spectra},\ }\href {https://doi.org/10.1063/1.4791712} {\bibfield
  {journal} {\bibinfo  {journal} {\jcp}\ }\textbf {\bibinfo {volume} {138}},\
  \bibinfo {pages} {084307} (\bibinfo {year} {2013})}\BibitemShut {NoStop}%
\bibitem [{\citenamefont {Chefdeville}\ \emph {et~al.}(2015)\citenamefont
  {Chefdeville}, \citenamefont {Stoecklin}, \citenamefont {Naulin},
  \citenamefont {Jankowski}, \citenamefont {Szalewicz}, \citenamefont {Faure},
  \citenamefont {Costes},\ and\ \citenamefont {Bergeat}}]{chefdeville15}%
  \BibitemOpen
  \bibfield  {author} {\bibinfo {author} {\bibfnamefont {S.}~\bibnamefont
  {Chefdeville}}, \bibinfo {author} {\bibfnamefont {T.}~\bibnamefont
  {Stoecklin}}, \bibinfo {author} {\bibfnamefont {C.}~\bibnamefont {Naulin}},
  \bibinfo {author} {\bibfnamefont {P.}~\bibnamefont {Jankowski}}, \bibinfo
  {author} {\bibfnamefont {K.}~\bibnamefont {Szalewicz}}, \bibinfo {author}
  {\bibfnamefont {A.}~\bibnamefont {Faure}}, \bibinfo {author} {\bibfnamefont
  {M.}~\bibnamefont {Costes}},\ and\ \bibinfo {author} {\bibfnamefont
  {A.}~\bibnamefont {Bergeat}},\ }\bibfield  {title} {\bibinfo {title}
  {Experimental and theoretical analysis of low-energy {CO} + {H$_2$} inelastic
  collisions},\ }\href {https://doi.org/10.1088/2041-8205/799/1/L9} {\bibfield
  {journal} {\bibinfo  {journal} {\apj}\ }\textbf {\bibinfo {volume} {799}},\
  \bibinfo {pages} {L9} (\bibinfo {year} {2015})}\BibitemShut {NoStop}%
\bibitem [{\citenamefont {Yang}\ \emph {et~al.}(2015)\citenamefont {Yang},
  \citenamefont {Zhang}, \citenamefont {Wang}, \citenamefont {Stancil},
  \citenamefont {Bowman}, \citenamefont {Balakrishnan},\ and\ \citenamefont
  {Forrey}}]{yang15}%
  \BibitemOpen
  \bibfield  {author} {\bibinfo {author} {\bibfnamefont {B.}~\bibnamefont
  {Yang}}, \bibinfo {author} {\bibfnamefont {P.}~\bibnamefont {Zhang}},
  \bibinfo {author} {\bibfnamefont {X.}~\bibnamefont {Wang}}, \bibinfo {author}
  {\bibfnamefont {P.~C.}\ \bibnamefont {Stancil}}, \bibinfo {author}
  {\bibfnamefont {J.~M.}\ \bibnamefont {Bowman}}, \bibinfo {author}
  {\bibfnamefont {N.}~\bibnamefont {Balakrishnan}},\ and\ \bibinfo {author}
  {\bibfnamefont {R.~C.}\ \bibnamefont {Forrey}},\ }\bibfield  {title}
  {\bibinfo {title} {Quantum dynamics of {CO--H$_2$} in full dimensionality},\
  }\href {https://doi.org/10.1038/ncomms7629} {\bibfield  {journal} {\bibinfo
  {journal} {Nat. Commun.}\ }\textbf {\bibinfo {volume} {6}},\ \bibinfo {pages}
  {6629} (\bibinfo {year} {2015})}\BibitemShut {NoStop}%
\bibitem [{\citenamefont {Yang}\ \emph {et~al.}(2016)\citenamefont {Yang},
  \citenamefont {Balakrishnan}, \citenamefont {Zhang}, \citenamefont {Wang},
  \citenamefont {Bowman}, \citenamefont {Forrey},\ and\ \citenamefont
  {Stancil}}]{yang16}%
  \BibitemOpen
  \bibfield  {author} {\bibinfo {author} {\bibfnamefont {B.~H.}\ \bibnamefont
  {Yang}}, \bibinfo {author} {\bibfnamefont {N.}~\bibnamefont {Balakrishnan}},
  \bibinfo {author} {\bibfnamefont {P.}~\bibnamefont {Zhang}}, \bibinfo
  {author} {\bibfnamefont {X.}~\bibnamefont {Wang}}, \bibinfo {author}
  {\bibfnamefont {J.~M.}\ \bibnamefont {Bowman}}, \bibinfo {author}
  {\bibfnamefont {R.~C.}\ \bibnamefont {Forrey}},\ and\ \bibinfo {author}
  {\bibfnamefont {P.~C.}\ \bibnamefont {Stancil}},\ }\bibfield  {title}
  {\bibinfo {title} {Full-dimensional quantum dynamics of {CO} in collision
  with {H$_2$}},\ }\href {https://doi.org/10.1063/1.4958951} {\bibfield
  {journal} {\bibinfo  {journal} {\jcp}\ }\textbf {\bibinfo {volume} {145}},\
  \bibinfo {pages} {034308} (\bibinfo {year} {2016})}\BibitemShut {NoStop}%
\bibitem [{\citenamefont {{Faure}}\ \emph {et~al.}(2016)\citenamefont
  {{Faure}}, \citenamefont {{Jankowski}}, \citenamefont {{Stoecklin}},\ and\
  \citenamefont {{Szalewicz}}}]{faure16}%
  \BibitemOpen
  \bibfield  {author} {\bibinfo {author} {\bibfnamefont {A.}~\bibnamefont
  {{Faure}}}, \bibinfo {author} {\bibfnamefont {P.}~\bibnamefont
  {{Jankowski}}}, \bibinfo {author} {\bibfnamefont {T.}~\bibnamefont
  {{Stoecklin}}},\ and\ \bibinfo {author} {\bibfnamefont {K.}~\bibnamefont
  {{Szalewicz}}},\ }\bibfield  {title} {\bibinfo {title} {{On the importance of
  full-dimensionality in low-energy molecular scattering calculations}},\
  }\href {https://doi.org/10.1038/srep28449} {\bibfield  {journal} {\bibinfo
  {journal} {Sci. Rep.}\ }\textbf {\bibinfo {volume} {6}},\ \bibinfo {eid}
  {28449} (\bibinfo {year} {2016})}\BibitemShut {NoStop}%
\bibitem [{\citenamefont {Castro}\ \emph {et~al.}(2017)\citenamefont {Castro},
  \citenamefont {Doan}, \citenamefont {Klemka}, \citenamefont {Forrey},
  \citenamefont {Yang}, \citenamefont {Stancil},\ and\ \citenamefont
  {Balakrishnan}}]{castro17}%
  \BibitemOpen
  \bibfield  {author} {\bibinfo {author} {\bibfnamefont {C.}~\bibnamefont
  {Castro}}, \bibinfo {author} {\bibfnamefont {K.}~\bibnamefont {Doan}},
  \bibinfo {author} {\bibfnamefont {M.}~\bibnamefont {Klemka}}, \bibinfo
  {author} {\bibfnamefont {R.}~\bibnamefont {Forrey}}, \bibinfo {author}
  {\bibfnamefont {B.~H.}\ \bibnamefont {Yang}}, \bibinfo {author}
  {\bibfnamefont {P.~C.}\ \bibnamefont {Stancil}},\ and\ \bibinfo {author}
  {\bibfnamefont {N.}~\bibnamefont {Balakrishnan}},\ }\bibfield  {title}
  {\bibinfo {title} {Inelastic cross sections and rate coefficients for
  collisions between co and h$_2$},\ }\href
  {https://doi.org/10.1016/j.molap.2017.01.003} {\bibfield  {journal} {\bibinfo
   {journal} {Mol. Astrophys.}\ }\textbf {\bibinfo {volume} {6}},\ \bibinfo
  {pages} {47} (\bibinfo {year} {2017})}\BibitemShut {NoStop}%
\bibitem [{\citenamefont {{Br{\'e}chignac}}\ \emph {et~al.}(1980)\citenamefont
  {{Br{\'e}chignac}}, \citenamefont {{Picard-Bersellini}}, \citenamefont
  {{Charneau}},\ and\ \citenamefont {{Launay}}}]{brechignac80}%
  \BibitemOpen
  \bibfield  {author} {\bibinfo {author} {\bibfnamefont {P.}~\bibnamefont
  {{Br{\'e}chignac}}}, \bibinfo {author} {\bibfnamefont {A.}~\bibnamefont
  {{Picard-Bersellini}}}, \bibinfo {author} {\bibfnamefont {R.}~\bibnamefont
  {{Charneau}}},\ and\ \bibinfo {author} {\bibfnamefont {J.~M.}\ \bibnamefont
  {{Launay}}},\ }\bibfield  {title} {\bibinfo {title} {Rotational relaxation of
  {CO} by collisions with {H$_{2}$} molecules: A comparison between theory and
  experiment},\ }\href {https://doi.org/10.1016/0301-0104(80)87065-0}
  {\bibfield  {journal} {\bibinfo  {journal} {Chem. Phys.}\ }\textbf {\bibinfo
  {volume} {53}},\ \bibinfo {pages} {165} (\bibinfo {year} {1980})}\BibitemShut
  {NoStop}%
\bibitem [{\citenamefont {{Stoecklin}}\ \emph {et~al.}(2017)\citenamefont
  {{Stoecklin}}, \citenamefont {{Faure}}, \citenamefont {{Jankowski}},
  \citenamefont {{Chefdeville}}, \citenamefont {{Bergeat}}, \citenamefont
  {{Naulin}}, \citenamefont {{Morales}},\ and\ \citenamefont
  {{Costes}}}]{stoecklin17}%
  \BibitemOpen
  \bibfield  {author} {\bibinfo {author} {\bibfnamefont {T.}~\bibnamefont
  {{Stoecklin}}}, \bibinfo {author} {\bibfnamefont {A.}~\bibnamefont
  {{Faure}}}, \bibinfo {author} {\bibfnamefont {P.}~\bibnamefont
  {{Jankowski}}}, \bibinfo {author} {\bibfnamefont {S.}~\bibnamefont
  {{Chefdeville}}}, \bibinfo {author} {\bibfnamefont {A.}~\bibnamefont
  {{Bergeat}}}, \bibinfo {author} {\bibfnamefont {C.}~\bibnamefont {{Naulin}}},
  \bibinfo {author} {\bibfnamefont {S.~B.}\ \bibnamefont {{Morales}}},\ and\
  \bibinfo {author} {\bibfnamefont {M.}~\bibnamefont {{Costes}}},\ }\bibfield
  {title} {\bibinfo {title} {Comparative experimental and theoretical study of
  the rotational excitation of {CO} by collision with {\em ortho}- and {\em
  para}-{D$_2$} molecules},\ }\href {https://doi.org/10.1039/C6CP06404C}
  {\bibfield  {journal} {\bibinfo  {journal} {Phys. Chem. Chem. Phys.}\
  }\textbf {\bibinfo {volume} {19}},\ \bibinfo {pages} {189} (\bibinfo {year}
  {2017})}\BibitemShut {NoStop}%
\bibitem [{\citenamefont {{Garberoglio}}\ \emph {et~al.}(2017)\citenamefont
  {{Garberoglio}}, \citenamefont {{Jankowski}}, \citenamefont {{Szalewicz}},\
  and\ \citenamefont {{Harvey}}}]{garberogli17}%
  \BibitemOpen
  \bibfield  {author} {\bibinfo {author} {\bibfnamefont {G.}~\bibnamefont
  {{Garberoglio}}}, \bibinfo {author} {\bibfnamefont {P.}~\bibnamefont
  {{Jankowski}}}, \bibinfo {author} {\bibfnamefont {K.}~\bibnamefont
  {{Szalewicz}}},\ and\ \bibinfo {author} {\bibfnamefont {A.~H.}\ \bibnamefont
  {{Harvey}}},\ }\bibfield  {title} {\bibinfo {title} {{All-dimensional
  H$_{2}$-CO potential: Validation with fully quantum second virial
  coefficients}},\ }\href {https://doi.org/10.1063/1.4974993} {\bibfield
  {journal} {\bibinfo  {journal} {\jcp}\ }\textbf {\bibinfo {volume} {146}},\
  \bibinfo {eid} {054304} (\bibinfo {year} {2017})}\BibitemShut {NoStop}%
\bibitem [{\citenamefont {{Rowe}}\ \emph {et~al.}(1984)\citenamefont {{Rowe}},
  \citenamefont {{Dupeyrat}}, \citenamefont {{Marquette}},\ and\ \citenamefont
  {{Gaucherel}}}]{CRESU}%
  \BibitemOpen
  \bibfield  {author} {\bibinfo {author} {\bibfnamefont {B.~R.}\ \bibnamefont
  {{Rowe}}}, \bibinfo {author} {\bibfnamefont {G.}~\bibnamefont {{Dupeyrat}}},
  \bibinfo {author} {\bibfnamefont {J.~B.}\ \bibnamefont {{Marquette}}},\ and\
  \bibinfo {author} {\bibfnamefont {P.}~\bibnamefont {{Gaucherel}}},\
  }\bibfield  {title} {\bibinfo {title} {Study of the reactions {N$^+_2$ +
  2N$_2$ → N$^+_4$ + N$_2$ and O$^+_2$ + 2O$_2$ → O$^+_4$ + O$_2$} from 20
  to 160 {K} by the {CRESU} technique},\ }\href
  {https://doi.org/10.1063/1.446513} {\bibfield  {journal} {\bibinfo  {journal}
  {\jcp}\ }\textbf {\bibinfo {volume} {80}},\ \bibinfo {pages} {4915} (\bibinfo
  {year} {1984})}\BibitemShut {NoStop}%
\bibitem [{\citenamefont {Sims}\ \emph {et~al.}(1994)\citenamefont {Sims},
  \citenamefont {Queffelec}, \citenamefont {Defrance}, \citenamefont
  {Rebrion-Rowe}, \citenamefont {Travers}, \citenamefont {Bocherel},
  \citenamefont {Rowe},\ and\ \citenamefont {Smith}}]{sims94}%
  \BibitemOpen
  \bibfield  {author} {\bibinfo {author} {\bibfnamefont {I.~R.}\ \bibnamefont
  {Sims}}, \bibinfo {author} {\bibfnamefont {J.~L.}\ \bibnamefont {Queffelec}},
  \bibinfo {author} {\bibfnamefont {A.}~\bibnamefont {Defrance}}, \bibinfo
  {author} {\bibfnamefont {C.}~\bibnamefont {Rebrion-Rowe}}, \bibinfo {author}
  {\bibfnamefont {D.}~\bibnamefont {Travers}}, \bibinfo {author} {\bibfnamefont
  {P.}~\bibnamefont {Bocherel}}, \bibinfo {author} {\bibfnamefont {B.~R.}\
  \bibnamefont {Rowe}},\ and\ \bibinfo {author} {\bibfnamefont {I.~W.~M.}\
  \bibnamefont {Smith}},\ }\bibfield  {title} {\bibinfo {title} {Ultralow
  temperature kinetics of neutral-neutral reactions - the technique and results
  for the reactions {CN} + {O$_2$} down to 13 {K} and {CN} + {NH$_3$} down to
  25 {K}},\ }\href {https://doi.org/10.1063/1.467227} {\bibfield  {journal}
  {\bibinfo  {journal} {\jcp}\ }\textbf {\bibinfo {volume} {100}},\ \bibinfo
  {pages} {4229} (\bibinfo {year} {1994})}\BibitemShut {NoStop}%
\bibitem [{\citenamefont {Tizniti}\ \emph {et~al.}(2014)\citenamefont
  {Tizniti}, \citenamefont {Le~Picard}, \citenamefont {Lique}, \citenamefont
  {Berteloite}, \citenamefont {Canosa}, \citenamefont {Alexander},\ and\
  \citenamefont {Sims}}]{Tizniti2014}%
  \BibitemOpen
  \bibfield  {author} {\bibinfo {author} {\bibfnamefont {M.}~\bibnamefont
  {Tizniti}}, \bibinfo {author} {\bibfnamefont {S.~D.}\ \bibnamefont
  {Le~Picard}}, \bibinfo {author} {\bibfnamefont {F.}~\bibnamefont {Lique}},
  \bibinfo {author} {\bibfnamefont {C.}~\bibnamefont {Berteloite}}, \bibinfo
  {author} {\bibfnamefont {A.}~\bibnamefont {Canosa}}, \bibinfo {author}
  {\bibfnamefont {M.~H.}\ \bibnamefont {Alexander}},\ and\ \bibinfo {author}
  {\bibfnamefont {I.~R.}\ \bibnamefont {Sims}},\ }\bibfield  {title} {\bibinfo
  {title} {Measurement of the rate of the {F + H$_2$} reaction at very low
  temperatures},\ }\href {https://doi.org/10.1038/nchem.1835} {\bibfield
  {journal} {\bibinfo  {journal} {Nat. Chem.}\ }\textbf {\bibinfo {volume}
  {6}},\ \bibinfo {pages} {141} (\bibinfo {year} {2014})}\BibitemShut {NoStop}%
\bibitem [{Sup()}]{Supplemental}%
  \BibitemOpen
  \href@noop {} {\bibinfo {title} {{See Supplemental Material at
  https://journals.aps.org/pra/ for further details about the IRVUVDR
  technique, low temperature uniform supersonic flows of hydrogen gas, and
  spectroscopic confirmation of uniform flow temperatures, as well as citations
  to Refs. \cite{Hilbig1983, Dupeyrat1985, Milenko1997, Mat2005,
  Western2017}.}}}\BibitemShut {Stop}%
\bibitem [{\citenamefont {Carty}\ \emph {et~al.}(2004)\citenamefont {Carty},
  \citenamefont {Goddard}, \citenamefont {Sims},\ and\ \citenamefont
  {Smith}}]{carty_rotational_2004}%
  \BibitemOpen
  \bibfield  {author} {\bibinfo {author} {\bibfnamefont {D.}~\bibnamefont
  {Carty}}, \bibinfo {author} {\bibfnamefont {A.}~\bibnamefont {Goddard}},
  \bibinfo {author} {\bibfnamefont {I.~R.}\ \bibnamefont {Sims}},\ and\
  \bibinfo {author} {\bibfnamefont {I.~W.~M.}\ \bibnamefont {Smith}},\
  }\bibfield  {title} {\bibinfo {title} {Rotational energy transfer in
  collisions between {CO}(${X}^1{\Sigma}^+, v=2, {J}=0, 1, 4$ and 6) and {He}
  at temperatures from 294 to 15 {K}},\ }\href
  {https://doi.org/10.1063/1.1780163} {\bibfield  {journal} {\bibinfo
  {journal} {\jcp}\ }\textbf {\bibinfo {volume} {121}},\ \bibinfo {pages}
  {4671} (\bibinfo {year} {2004})}\BibitemShut {NoStop}%
\bibitem [{\citenamefont {Mertens}\ \emph {et~al.}(2017)\citenamefont
  {Mertens}, \citenamefont {Labiad}, \citenamefont {Denis-Alpizar},
  \citenamefont {Fournier}, \citenamefont {Carty}, \citenamefont {Picard},
  \citenamefont {Stoecklin},\ and\ \citenamefont {Sims}}]{mertens_rotational}%
  \BibitemOpen
  \bibfield  {author} {\bibinfo {author} {\bibfnamefont {L.~A.}\ \bibnamefont
  {Mertens}}, \bibinfo {author} {\bibfnamefont {H.}~\bibnamefont {Labiad}},
  \bibinfo {author} {\bibfnamefont {O.}~\bibnamefont {Denis-Alpizar}}, \bibinfo
  {author} {\bibfnamefont {M.}~\bibnamefont {Fournier}}, \bibinfo {author}
  {\bibfnamefont {D.}~\bibnamefont {Carty}}, \bibinfo {author} {\bibfnamefont
  {S.~D.~L.}\ \bibnamefont {Picard}}, \bibinfo {author} {\bibfnamefont
  {T.}~\bibnamefont {Stoecklin}},\ and\ \bibinfo {author} {\bibfnamefont
  {I.~R.}\ \bibnamefont {Sims}},\ }\bibfield  {title} {\bibinfo {title}
  {Rotational energy transfer in collisions between {CO} and {Ar} at
  temperatures from 293 to 30 {K}},\ }\href
  {https://doi.org/10.1016/j.cplett.2017.05.052} {\bibfield  {journal}
  {\bibinfo  {journal} {Chem. Phys. Lett.}\ }\textbf {\bibinfo {volume}
  {683}},\ \bibinfo {pages} {521} (\bibinfo {year} {2017})}\BibitemShut
  {NoStop}%
\bibitem [{\citenamefont {James}\ \emph {et~al.}(1998)\citenamefont {James},
  \citenamefont {Sims}, \citenamefont {Smith}, \citenamefont {Alexander},\ and\
  \citenamefont {Yang}}]{james_combined_1998}%
  \BibitemOpen
  \bibfield  {author} {\bibinfo {author} {\bibfnamefont {P.~L.}\ \bibnamefont
  {James}}, \bibinfo {author} {\bibfnamefont {I.~R.}\ \bibnamefont {Sims}},
  \bibinfo {author} {\bibfnamefont {I.~W.~M.}\ \bibnamefont {Smith}}, \bibinfo
  {author} {\bibfnamefont {M.~H.}\ \bibnamefont {Alexander}},\ and\ \bibinfo
  {author} {\bibfnamefont {M.}~\bibnamefont {Yang}},\ }\bibfield  {title}
  {\bibinfo {title} {A combined experimental and theoretical study of
  rotational energy transfer in collisions between {NO}(${X}^1{\Pi}_{1/2},
  v=3,{J}$) and {He}, {Ar} and {N}$_2$ at temperatures down to 7 {K}},\ }\href
  {https://doi.org/10.1063/1.476517} {\bibfield  {journal} {\bibinfo  {journal}
  {\jcp}\ }\textbf {\bibinfo {volume} {109}},\ \bibinfo {pages} {3882}
  (\bibinfo {year} {1998})}\BibitemShut {NoStop}%
\bibitem [{\citenamefont {{Hutson}}\ and\ \citenamefont
  {{Green}}(2012)}]{molscat}%
  \BibitemOpen
  \bibfield  {author} {\bibinfo {author} {\bibfnamefont {J.~M.}\ \bibnamefont
  {{Hutson}}}\ and\ \bibinfo {author} {\bibfnamefont {S.}~\bibnamefont
  {{Green}}},\ }\href@noop {} {\bibinfo {title} {{MOLSCAT: MOLecular
  SCATtering}}},\ \bibinfo {howpublished} {Astrophysics Source Code Library}
  (\bibinfo {year} {2012}),\ \Eprint {https://arxiv.org/abs/1206.004}
  {ascl:1206.004} \BibitemShut {NoStop}%
\bibitem [{\citenamefont {Huber}\ and\ \citenamefont
  {Herzberg}(1979)}]{Huber:79}%
  \BibitemOpen
  \bibfield  {author} {\bibinfo {author} {\bibfnamefont {K.~P.}\ \bibnamefont
  {Huber}}\ and\ \bibinfo {author} {\bibfnamefont {G.}~\bibnamefont
  {Herzberg}},\ }\href@noop {} {\emph {\bibinfo {title} {Molecular Spectra and
  Molecular Structure. {IV}. {C}onstants of diatomic molecules}}}\ (\bibinfo
  {publisher} {Van Nostrand Reinhold},\ \bibinfo {address} {New York},\
  \bibinfo {year} {1979})\BibitemShut {NoStop}%
\bibitem [{\citenamefont {Mahan}(1975)}]{Mahan1974}%
  \BibitemOpen
  \bibfield  {author} {\bibinfo {author} {\bibfnamefont {B.~H.}\ \bibnamefont
  {Mahan}},\ }\bibfield  {title} {\bibinfo {title} {Microscopic reversibility
  and detailed balance. an analysis},\ }\href
  {https://doi.org/10.1021/ed052p299} {\bibfield  {journal} {\bibinfo
  {journal} {J. Chem. Educ.}\ }\textbf {\bibinfo {volume} {52}},\ \bibinfo
  {pages} {299} (\bibinfo {year} {1975})}\BibitemShut {NoStop}%
\bibitem [{\citenamefont {McCurdy}\ and\ \citenamefont
  {Miller}(1977)}]{mccurdy77}%
  \BibitemOpen
  \bibfield  {author} {\bibinfo {author} {\bibfnamefont {C.~W.}\ \bibnamefont
  {McCurdy}}\ and\ \bibinfo {author} {\bibfnamefont {W.~H.}\ \bibnamefont
  {Miller}},\ }\bibfield  {title} {\bibinfo {title} {Interference effects in
  rotational state distributions - propensity and inverse propensity},\ }\href
  {https://doi.org/10.1063/1.434890} {\bibfield  {journal} {\bibinfo  {journal}
  {\jcp}\ }\textbf {\bibinfo {volume} {67}},\ \bibinfo {pages} {463} (\bibinfo
  {year} {1977})}\BibitemShut {NoStop}%
\bibitem [{\citenamefont {Hilbig}\ and\ \citenamefont
  {Wallenstein}(1983)}]{Hilbig1983}%
  \BibitemOpen
  \bibfield  {author} {\bibinfo {author} {\bibfnamefont {R.}~\bibnamefont
  {Hilbig}}\ and\ \bibinfo {author} {\bibfnamefont {R.}~\bibnamefont
  {Wallenstein}},\ }\bibfield  {title} {\bibinfo {title} {Tunable {VUV}
  radiation generated by two-photon resonant frequency mixing in xenon},\
  }\href {https://doi.org/10.1109/jqe.1983.1071833} {\bibfield  {journal}
  {\bibinfo  {journal} {{IEEE} J. Quantum. Electron.}\ }\textbf {\bibinfo
  {volume} {19}},\ \bibinfo {pages} {194} (\bibinfo {year} {1983})}\BibitemShut
  {NoStop}%
\bibitem [{\citenamefont {Dupeyrat}\ \emph {et~al.}(1985)\citenamefont
  {Dupeyrat}, \citenamefont {Marquette},\ and\ \citenamefont
  {Rowe}}]{Dupeyrat1985}%
  \BibitemOpen
  \bibfield  {author} {\bibinfo {author} {\bibfnamefont {G.}~\bibnamefont
  {Dupeyrat}}, \bibinfo {author} {\bibfnamefont {J.~B.}\ \bibnamefont
  {Marquette}},\ and\ \bibinfo {author} {\bibfnamefont {B.~R.}\ \bibnamefont
  {Rowe}},\ }\bibfield  {title} {\bibinfo {title} {Design and testing of
  axisymmetric nozzles for ion-molecule reaction studies between
  20{\hspace{0.167em}}{\textdegree}{K} and
  160{\hspace{0.167em}}{\textdegree}{K}},\ }\href
  {https://doi.org/10.1063/1.865010} {\bibfield  {journal} {\bibinfo  {journal}
  {Phys. Fluids}\ }\textbf {\bibinfo {volume} {28}},\ \bibinfo {pages} {1273}
  (\bibinfo {year} {1985})}\BibitemShut {NoStop}%
\bibitem [{\citenamefont {Milenko}\ \emph {et~al.}(1997)\citenamefont
  {Milenko}, \citenamefont {Sibileva},\ and\ \citenamefont
  {Strzhemechny}}]{Milenko1997}%
  \BibitemOpen
  \bibfield  {author} {\bibinfo {author} {\bibfnamefont {Y.~Y.}\ \bibnamefont
  {Milenko}}, \bibinfo {author} {\bibfnamefont {R.~M.}\ \bibnamefont
  {Sibileva}},\ and\ \bibinfo {author} {\bibfnamefont {M.~A.}\ \bibnamefont
  {Strzhemechny}},\ }\bibfield  {title} {\bibinfo {title} {Natural ortho-para
  conversion rate in liquid and gaseous hydrogen},\ }\href
  {https://doi.org/10.1007/bf02396837} {\bibfield  {journal} {\bibinfo
  {journal} {J. Low Temp. Phys.}\ }\textbf {\bibinfo {volume} {107}},\ \bibinfo
  {pages} {77} (\bibinfo {year} {1997})}\BibitemShut {NoStop}%
\bibitem [{\citenamefont {Mat{\'{e}}}\ \emph {et~al.}(2005)\citenamefont
  {Mat{\'{e}}}, \citenamefont {Thibault}, \citenamefont {Tejeda}, \citenamefont
  {Fern{\'{a}}ndez},\ and\ \citenamefont {Montero}}]{Mat2005}%
  \BibitemOpen
  \bibfield  {author} {\bibinfo {author} {\bibfnamefont {B.}~\bibnamefont
  {Mat{\'{e}}}}, \bibinfo {author} {\bibfnamefont {F.}~\bibnamefont
  {Thibault}}, \bibinfo {author} {\bibfnamefont {G.}~\bibnamefont {Tejeda}},
  \bibinfo {author} {\bibfnamefont {J.~M.}\ \bibnamefont {Fern{\'{a}}ndez}},\
  and\ \bibinfo {author} {\bibfnamefont {S.}~\bibnamefont {Montero}},\
  }\bibfield  {title} {\bibinfo {title} {Inelastic collisions in para-h2:
  Translation-rotation state-to-state rate coefficients and cross sections at
  low temperature and energy},\ }\href {https://doi.org/10.1063/1.1850464}
  {\bibfield  {journal} {\bibinfo  {journal} {\jcp}\ }\textbf {\bibinfo
  {volume} {122}},\ \bibinfo {pages} {064313} (\bibinfo {year}
  {2005})}\BibitemShut {NoStop}%
\bibitem [{\citenamefont {Western}(2017)}]{Western2017}%
  \BibitemOpen
  \bibfield  {author} {\bibinfo {author} {\bibfnamefont {C.~M.}\ \bibnamefont
  {Western}},\ }\bibfield  {title} {\bibinfo {title} {{PGOPHER}: A program for
  simulating rotational, vibrational and electronic spectra},\ }\href
  {https://doi.org/10.1016/j.jqsrt.2016.04.010} {\bibfield  {journal} {\bibinfo
   {journal} {J. Quant. Spectrosc. Radiat. Transfer}\ }\textbf {\bibinfo
  {volume} {186}},\ \bibinfo {pages} {221} (\bibinfo {year}
  {2017})}\BibitemShut {NoStop}%
\end{thebibliography}%
\end{document}